\documentclass[12pt, prd, showpacs]{revtex4}
\usepackage{amssymb}
\usepackage{amsmath}

\input{tcilatex}

\begin{document}

\title{Center of mass energy of colliding electrically neutral particles and
super-Penrose process }
\author{O. B. Zaslavskii}
\affiliation{Department of Physics and Technology, Kharkov V.N. Karazin National
University, 4 Svoboda Square, Kharkov 61022, Ukraine}
\affiliation{Institute of Mathematics and Mechanics, Kazan Federal University, 18
Kremlyovskaya St., Kazan 420008, Russia}
\email{zaslav@ukr.net }

\begin{abstract}
We consider collisions of electrically neutral particles in the background
of axially symmetric rotating space-times. For reactions of the kind 1+2$%
\rightarrow $3+4 we give full classification of possible scenarios based on
exact expressions for dynamic characteristics of particles 3 and 4 depending
on particles 1 and 2. There are five nonequivalent scenarios valid for
different types of space-time (black hole, naked singularity, etc.). We are
mainly interested in the situations when particle collisions can give
unbounded outcome for (i) the energy $E_{c.m.}$ in the centre of mass frame
and (ii) Killing energy $E$ at infinity. If (i) is fulfilled, this may or
may not lead to (ii). If (ii)\ holds, this is called the super-Penrose
process (SPP). For equatorial particle motion, we establish close relation
between the type of behavior of $E_{c.m.}$ depending on the lapse function $%
N $ in the point of collision and the possibility of the SPP. This unites
separate previous observations in literature in a unified picture as a
whole. In doing so, no explicit transformations to the centre of mass frame
and back are needed, so all consideration is carried out in the original
frame. As a result,we suggest classification of sub-classes of scenarios
that are able (or unable) to give rise to the SPP particle collision,
super-Penrose process, centre of mass frame.
\end{abstract}

\maketitle

\section{Introduction}

In recent years, a possibility of black holes to serve as supercolliders
attracts attention after findings by Ba\~{n}ados, Silk and West (the BSW
effect, after the names of its authors) \cite{ban} who showed that under
some conditions, collision between two particles near the extremal Kerr
metric leads to the unbounded energy $E_{c.m.}$ in the centre of mass (CM)
frame. As a result, interest to previous findings on high-energy collisions
near black holes \cite{pir1} - \cite{pir3} also revived. This effect was
extended to rather generic black holes including nonextremal ones.
Meanwhile, especially interesting is a question about the energy $E$
measured by an observer at infinity since it would give a chance to register
debris of corresponding collision in the laboratory, at least in principle.
If $E$ can be unbounded, this is called the super-Penrose process (SPP).
However, detailed investigations showed that rotating neutral black holes
are not pertinent for this purpose in the BSW scenario, so in spite of
unbounded $E_{c.m.}$, the energy $E$ gained in the BSW process, remains
quite modest \cite{j} - \cite{z}. Another type of scenario in which a
fine-tuned outgoing particle experiences head-on collision with an incoming
one increases the efficiency of the process \cite{shnit} but, anyway, it
remains bounded \cite{cons} - \cite{frac}.

There have been attempts to extend the results of \cite{shnit} to the case
when both particles are generic (so-called usual), not fine-tuned.
Numerically, it was observed in \cite{card} that such a collision near the
horizon formally leads to the SPP. This was found independently in \cite{mod}
analytically for rather generic black holes. The escape probability for such
processes was found analytically for the Kerr black hole \cite{esc}. There
is a problem, however, that \ such a usual particle with a finite energy
cannot emerge from a black hole \cite{frac} although this is possible for
white holes \cite{gpw}. Examination of more involved scenarios of multiple
scattering showed that this is also impossible for black holes, provided all
characteristics of initial ingoing articles are finite \cite{pir15}, \cite%
{epl}. Thus, one is led to reject black holes as potential supercolliders of
particles coming to infinity (although near the horizon this is quite
possible).

From the other hand, there are scenarios of collisions in the background of
naked singularities and wormholes \cite{headon} - \cite{owh} when the SPP\
does occur. Thus there is a series of particular observations for some kinds
of physical objects with a positive result in this sense.

The question arises, what is the \ underlying reason, why the SPP is
possible in the aforementioned case but is forbidden for black holes? In the
present work we consider rotating neutral space-times and suggest
explanation. We argue that so different behavior follows unambiguously from
one feature - the dependence of $E_{c.m.}$ on the lapse function $N$ in the
point of collision. Thus we relate key properties of the effect under
consideration in the region of strong gravity (say, near the horizon of a
black hole) and at infinity. At the same time, we present simple and exact
formulas for collisions in the equatorial plane that can be useful in a
quite general context for other purposes as well. This enabled us to
classify all relevant scenarios of collisions in the equatorial plane
including those pertinent to the SPP. Thus we give a unified picture as a
whole from which previous observations about (im)possibility of the SPP
follow as particular cases.

We would like to stress that in the present paper we consider collisions of
electrically neutral particles only. Experience shows that for the
electrically charge case \cite{rn} the results can be very different from
the neutral one described above.

We use the geometric system of units in which fundamental constants $G=c=1$.

\section{Basic formulas}

Let us consider the metric%
\begin{equation}
ds^{2}=-N^{2}dt^{2}+g_{\phi }(d\phi -\omega dt)^{2}+\frac{dr^{2}}{A}+g_{\phi
}d\theta ^{2}\text{,}
\end{equation}%
where the coefficients do not depend on $t$ and $\phi $. (To simplify
formulas, we use notation $g_{\phi }$ for the component of the metric tensor 
$g_{\phi \phi }$). We suppose that the equatorial plane is a plane of
symmetry and are interested in the motion within this plane only. On the
horizon $r=r_{+}$ we have $N=0$. Equations of motion read%
\begin{equation}
m\dot{t}=\frac{X}{N^{2}}\text{,}  \label{mt}
\end{equation}%
\begin{equation}
m\frac{N}{\sqrt{A}}\dot{r}=P_{r}=\sigma P\text{,}
\end{equation}%
\begin{equation}
m\dot{\phi}=\frac{L}{g_{\phi }}+\frac{\omega X}{N^{2}}\text{, }  \label{phi}
\end{equation}%
where dot denotes differentiation with respect to the proper time $\tau $, 
\begin{equation}
X=E-\omega L\text{,}  \label{x}
\end{equation}%
$E$ being the conserved energy, $L\,\ $conserved angular momentum, $\sigma
=\pm 1$ depending on the direction of motion.%
\begin{equation}
P=\sqrt{X^{2}-N^{2}\tilde{m}^{2}},\text{ }  \label{zn}
\end{equation}%
\begin{equation}
\tilde{m}^{2}=m^{2}+\frac{L^{2}}{g_{\phi }}\text{.}  \label{mm}
\end{equation}

The forward-in-time condition $\dot{t}>0$ entails%
\begin{equation}
X\geq 0\text{.}  \label{ft}
\end{equation}

\section{Collision vs decay}

Our main goal consists in the analysis of particle collisions. Particle 1
and 2 collide to produce particles 3 and 4. To simplify formulas, we
conditionally represent collision of particles 1 and 2 as creation of some
effective particle 0 that decays immediately to particles 3 and 4.
Hereafter, we use subscript $i$ ($i=1,2,3,4)$ to enumerate characteristics
of particles ($E_{i}$, $L_{i}$, $m_{i}$, $P_{i}$, etc.). Then, the energy
and angular momentum are given by the conservation laws:

\begin{equation}
E_{0}=E_{1}+E_{2}=E_{3}+E_{4}\text{,}
\end{equation}%
\begin{equation}
L_{0}=L_{1}+L_{2}=L_{3}+L_{4}\text{,}  \label{L0}
\end{equation}%
As a consequence, we have also the equality%
\begin{equation}
X_{0}=X_{1}+X_{2}=X_{3}+X_{4}\text{.}  \label{x0}
\end{equation}%
It is assumed that $X_{0}$ is some finite quantity. Then, as (\ref{ft}) is
valid for each particle, $X_{3}$ and $X_{4}$ are also bounded.

We also assume the initial radial momentum to be negative (particle 0 moves
towards a black hole), so particles 3 and 4 cannot have $P_{r}>0$ both. For
definiteness, we assume that it is particle 4 that moves inside always, so $%
\sigma _{4}=-1$. The conservation of the radial momentum in the reaction $%
1+2\rightarrow 0$ reads 
\begin{equation}
-P_{0}=\sigma _{2}P_{2}+\sigma _{1}P_{1}.  \label{p}
\end{equation}
In a similar way, for decay $0\rightarrow 3+4$ we have%
\begin{equation}
-P_{0}=\sigma _{3}P_{3}-P_{4}.  \label{p34}
\end{equation}

Taking the square of (\ref{p}), one can check that%
\begin{equation}
m_{0}^{2}=E_{c.m.}^{2},  \label{m0}
\end{equation}%
where $E_{c.m.}$ is the energy in the CM frame,%
\begin{equation}
E_{c.m.}^{2}=-(p_{1\mu }+p_{2\mu })(p_{1}^{\mu }+p_{2}^{\mu })\text{,}
\label{cmg}
\end{equation}%
$p_{i}^{\mu }=m_{i}u_{i}^{\mu }$, where $u_{i}^{\mu }$ is the four-velocity
of particle i. It follows from (\ref{cmg}) that 
\begin{equation}
E_{c.m.}^{2}=m_{1}^{2}+m_{2}^{2}+2m_{1}m_{2}\gamma ,  \label{cm}
\end{equation}%
the Lorentz factor of relative motion%
\begin{equation}
\gamma =-u_{1\mu }u_{2}^{\mu }\text{.}
\end{equation}%
In terms of characteristics of particles 1 and 2, one obtains from (\ref{mt}%
) - (\ref{phi}) that%
\begin{equation}
\gamma m_{1}m_{2}=\left( \frac{X_{1}X_{2}-\sigma _{1}\sigma _{2}P_{1}P_{2}}{%
N^{2}}-\frac{L_{1}L_{2}}{g_{\phi }}\right) _{c}\text{.}  \label{ga}
\end{equation}

Hereafter, subscript "c" indicates the point of collision. The similar
formulas hold if we calculate $E_{c.m.}$ between particles 3 and 4. Then, $%
E_{c.m.}$ is the same due to the conservation laws.

\section{Explicit solutions of equations}

One can solve eq. (\ref{p34}) exactly, taking into account (\ref{zn}). Then,
one finds

\begin{equation}
\left( X_{3}\right) _{c}=\frac{1}{2\tilde{m}_{0}^{2}}(X_{0}\Delta _{+}+P_{0}%
\sqrt{D}\delta )_{c}\text{,}  \label{x1}
\end{equation}%
\begin{equation}
\left( X_{4}\right) _{c}=\frac{1}{2\tilde{m}_{0}^{2}}(X_{0}\Delta _{-}-P_{0}%
\sqrt{D}\delta )_{c}\text{,}  \label{x2}
\end{equation}%
where $\delta =1$ or $\delta =-1$. 
\begin{equation}
\Delta _{\pm }=\tilde{m}_{0}^{2}\pm (\tilde{m}_{3}^{2}-\tilde{m}_{4}^{2}).
\label{del}
\end{equation}%
The positivity of $X_{3.4}$ entails 
\begin{equation}
\Delta _{\pm }>0.  \label{dpos}
\end{equation}%
\begin{equation}
D=\Delta _{+}^{2}-4\tilde{m}_{0}^{2}\tilde{m}_{3}^{2}=\Delta _{-}^{2}-4%
\tilde{m}_{0}^{2}\tilde{m}_{4}^{2}.  \label{D}
\end{equation}%
It is necessary that%
\begin{equation}
D\geq 0.  \label{d}
\end{equation}%
Then, after simple algebraic manipulations, (\ref{del}) - (\ref{d}) lead to%
\begin{equation}
\tilde{m}_{0}\geq \tilde{m}_{3}+\tilde{m}_{4}\text{.}  \label{m034}
\end{equation}

In the above formulas, all quantities related to particles 1 and 2 (hence,
those of effective particle 0 as well) are fixed. We also assume that masses 
$m_{3,4}$ are fixed for any given process. Meanwhile, one of two angular
momentum (say, $L_{3}$) remains a free parameter.

Then, by substitution into (\ref{zn}), one finds 
\begin{equation}
P_{3}=\frac{\left\vert P_{0}\Delta _{+}+\delta X_{0}\sqrt{D}\right\vert }{2%
\tilde{m}_{0}^{2}},  \label{p1}
\end{equation}%
\ \ \ \ \ \ 
\begin{equation}
P_{4}=\frac{\left\vert P_{0}\Delta _{-}-\delta X_{0}\sqrt{D}\right\vert }{2%
\tilde{m}_{0}^{2}},  \label{p2}
\end{equation}%
where we omitted subscript "c" for shortness. It is easy to check that $%
X_{i}^{2}-P_{i}^{2}=\tilde{m}_{i}^{2}N^{2}$ as it should be. The difference

\begin{equation}
H_{3}\equiv \frac{X_{0}^{2}D-P_{0}^{2}\Delta _{+}^{2}}{\tilde{m}_{0}^{2}}%
=\Delta _{+}^{2}N^{2}-4\tilde{m}_{3}^{2}X_{0}^{2}  \label{h3}
\end{equation}%
shows which term dominates the numerator in (\ref{p1}). In a similar way, $%
H_{4}$ is defined, with $\Delta _{+}$ replaced with $\Delta _{-}$ and $%
\tilde{m}_{3}$ replaced with $\tilde{m}_{4}$. Correspondingly, it is
convenient to introduce the factors $\varepsilon _{i}$ according to%
\begin{equation}
\varepsilon _{i}=sgnH_{i}\text{, }  \label{eps}
\end{equation}%
\newline
$i=3,4$.

Then, we can characterize the process by the set $(\sigma _{3},\varepsilon
_{3},\varepsilon _{4},\delta )$. Direct inspection by substitution of (\ref%
{p1}), (\ref{p2}) in (\ref{p34}) shows that the list of possible scenarios
includes six cases: 
\begin{equation}
1(+,+,+,-),2(+,+,-,-),3(-,-,-,+),4(-,+,-,+),5(-,-,-,-),6(-,-,+,-).
\label{list}
\end{equation}%
Cases 3 and 5 are equivalent. They differ by particle labels 3 and 4 only.
Thus actual number of different scenarios is five. In turning points, $%
H_{i}=0$ and the difference between $\varepsilon _{i}=+1$ and $\varepsilon
_{i}=-1$ becomes irrelevant.

Below, we examine three typical cases depending on the behavior of $%
E_{c.m.}=m_{0}$ and the region where collision occurs. Our main goal is to
establish, when the SPP is possible. In all cases we assume that all
characteristics ($m,E,L$) of initial particles 1 and 2 are finite.

To classify particles, in what follows we use the standard terminology but
in somewhat extended manner. Namely, we call a particle usual if $%
\lim_{N_{c}\rightarrow 0}X\neq 0$ and critical if $\lim_{N_{c}\rightarrow
0}X=0$. For black holes, this coincides with the standard definition \cite%
{ban}, \cite{prd}. But it includes also the case of naked singularities and
wormholes if collision occurs in the point with small but nonzero $%
N_{c}>\left( N_{c}\right) _{\min }$ when $\left( N_{c}\right) _{\min
}\rightarrow 0$ $\ $and $N_{c}\rightarrow 0$ \cite{owh}.

\section{Energy in CM frame is finite}

Let us consider collisions with finite $m_{0}$. It was observed in \cite{inf}
for the overspinned Kerr metric that if $m_{0}$ is finite, no SPP occurs.
This was generalized and explained in terms of the Wald identities \cite%
{wald} in \cite{waldcol}. In this sense, the absence of the SPP is clear in
advance. However, in this Section we reconsider this issue again to
demonstrate how our approach works. Collisions under discussion can be
realized in two typical situations: (i) in some intermediate point with $%
N_{c}=O(1),$ (ii) for two usual particles near the horizon, where $%
N\rightarrow 0$ \cite{ban}, \cite{prd}. Both $X_{3}$ and $X_{4}$ are finite
and nonzero. As $X_{3}$ should be finite, the property $E_{3}\rightarrow
\infty $ requires $L_{3}\rightarrow \infty $ as well to compensate big $%
E_{3} $. In doing so, $L_{4}=L_{0}-L_{3}\rightarrow -\infty $. As a result,
both $\tilde{m}_{3}~$and $\tilde{m}_{4}\rightarrow \infty $ according to (%
\ref{mm}). As $\tilde{m}_{0}$ remains finite by assumption, we see that eq. (%
\ref{m034}) is violated and so is condition (\ref{d}). As a result, the
scenario under discussion is impossible, so there is no SPP in this case, as
expected.

\section{Square of Energy in CM frame growing as $N_{c}^{-2}$}

Now, we consider collisions with $N_{c}\rightarrow 0$ and%
\begin{equation}
m_{0}^{2}\approx \frac{\alpha }{N_{c}^{2}}\text{, }N_{c}\rightarrow 0\text{.}
\label{n2}
\end{equation}

This can be realized if two particles experience head-on collision, so $%
\sigma _{1}\sigma _{2}=-1$ in (\ref{ga}). There are different objects and
scenarios with these features. Particle 1 can bounce back from the potential
barrier existing in the case of naked singularity and collide with an
ingoing particle 2 \cite{headon}, \cite{inf}, \cite{waldcol}. Then,
considering reflection from the potential barrier and making transformation
between the original stationary frame, LNRF (locally nonrotating frame) and
the CM one, we obtain for generic rotating axially symmetric over-spinned
space-time that $E_{3}$ can be unbounded, so the SPP does occur - see Sec.
3.5 of \cite{waldcol}. In another version, particles 1 and 2 can come from
the opposite mouths of a wormhole \cite{kras}, \cite{owh}.

Now, we are in position to show that the dependence (\ref{n2}) itself leads
to the SPP, independently of the details of the scenario. In doing so, our
approach is simpler than in \cite{waldcol} since we are using one frame only.

It follows from (\ref{cm}), (\ref{ga}) that 
\begin{equation}
\alpha =4(X_{1}X_{2})_{c}\approx 4(X_{1}X_{2})_{N=0}\text{.}  \label{a1}
\end{equation}

Then, we can again take the limit $L_{3}\rightarrow \infty $. But now, there
is some additional restriction to guarantee the finiteness of $X_{3}$, $%
X_{4} $ and the condition that, according to (\ref{zn}), 
\begin{equation}
X_{3,4}\geq N\tilde{m}_{3,4}\text{.}  \label{esc}
\end{equation}%
We consider at first the case when the centrifugal contribution to
\thinspace $P_{3,4}$ (\ref{zn}) due to $L_{3,4}^{2}$ has the same order as $%
X_{3,4}^{2}$. To this end, we assume that $L_{3}$ is adjusted to the
position of collision according to%
\begin{equation}
L_{3}\approx \frac{l}{N_{c}}\text{,}  \label{Ll}
\end{equation}%
where $l$ is some constant.

From (\ref{n2}), we have 
\begin{equation}
\tilde{m}_{0}^{2}\approx m_{0}^{2}\approx \frac{\alpha }{N_{c}^{2}},
\label{m0a}
\end{equation}%
since $L_{0}$ entering (\ref{mm}) is supposed to be \ finite. Meanwhile,%
\begin{equation}
\tilde{m}_{3}^{2}\approx \tilde{m}_{4}^{2}\approx \frac{b^{2}}{N_{c}^{2}}%
\text{,}  \label{m3a}
\end{equation}%
where%
\begin{equation}
b\equiv \frac{l}{\sqrt{\left( g_{\phi }\right) _{c}}}\text{,}  \label{b}
\end{equation}%
\begin{equation}
\Delta _{\pm }\approx \tilde{m}_{0}^{2}+O(\frac{1}{N})\text{, }D\approx 
\tilde{m}_{0}^{2}(\tilde{m}_{0}^{2}-4\tilde{m}_{3}^{2})=\frac{d^{2}}{%
N_{c}^{4}}\text{, }d=\sqrt{\alpha (\alpha -4b^{2})}\text{. }  \label{dela}
\end{equation}%
We must require 
\begin{equation}
b^{2}\leq \frac{\alpha }{4}\text{,}  \label{lmax}
\end{equation}%
so $l^{2}\leq \frac{\alpha }{4}\left( g_{\phi }\right) _{c}$. This agrees
with (\ref{m034}). Then, we get from (\ref{x1}), (\ref{x2})%
\begin{equation}
\left( X_{3,4}\right) _{c}\approx \frac{1}{2}[\left( X_{0}\right) _{c}\pm
\left( P_{0}\right) _{c}\frac{d}{\alpha }]\text{.}  \label{x34}
\end{equation}%
Here, for $N_{c}\rightarrow 0$, 
\begin{equation}
\left( P_{0}\right) _{c}\approx \sqrt{\left( X_{0}^{2}\right) _{c}-\alpha }%
=\left\vert X_{1}-X_{2}\right\vert _{c}  \label{pxa}
\end{equation}%
according to (\ref{zn}) and (\ref{a1}). The procedure is self-consistent, so
we can achieve $E_{3}\rightarrow \infty $, if we take $N_{c}$ sufficiently
small, with $L_{3}$ given by (\ref{Ll}). This hints at the SPP, provided the
escape condition is satisfied (see below).

To identify a type of scenario, we must evaluate (\ref{h3}):%
\begin{equation}
H_{3}\approx \left( \alpha ^{2}-4b^{2}X_{0}^{2}\right) N_{c}^{-2}\approx
H_{4}\text{.}
\end{equation}

We want to realize scenario 1 from list (\ref{list}) to have particle 3
escaping directly to infinity. Otherwise, this usual particle would fall in
a black hole. Then, we must require, in addition to (\ref{lmax}), also%
\begin{equation}
b^{2}\leq \frac{\alpha ^{2}}{4X_{0}^{2}}\text{.}  \label{l2}
\end{equation}%
It is seen from (\ref{x0}) and (\ref{a1}) that 
\begin{equation}
\alpha <X_{0}^{2},  \label{ax}
\end{equation}%
so the the expression inside the equare root in (\ref{pxa}) is indeed
non-negative, as it should be, and condition (\ref{l2}) is more tight that (%
\ref{lmax}).

This is not the end of story since the desired conditions near the point of
collision do not guarantee escaping. For escape to occur, we must require (%
\ref{esc}) in any point, not only in the point of collision. Bearing in mind
(\ref{x}), we can write (\ref{esc}) as%
\begin{equation}
\left( X_{3}\right) _{c}+L_{3}(\omega _{c}-\omega )>N\tilde{m}_{3}\text{.}
\label{l3}
\end{equation}

For large $L_{3}>0$, it is seen from (\ref{mm}) that%
\begin{equation}
\tilde{m}_{3}\approx \frac{L_{3}}{\sqrt{g_{\phi }}}\text{.}
\end{equation}

Then, this condition \ translates to%
\begin{equation}
Y\equiv \left( X_{3}\right) _{c}+L_{3}(\omega _{c}-\omega _{+})>0\text{, }%
\omega _{+}\equiv \omega +\frac{N}{\sqrt{g_{\phi }}}.  \label{Y}
\end{equation}%
The quantity $\omega _{+}$ has a simple physical meaning. As is known, it
defines the maximum angular velocity of a particle in the ergoregion.

We want to have inequality $Y_{c}=\left( X_{3}\right) _{c}-\frac{l}{\sqrt{%
\left( g_{\phi }\right) _{c}}}=\left( X_{3}\right) _{c}-b>0$. Taking into
account (\ref{dela}), (\ref{x34})\ we can write%
\begin{equation}
\left( X_{0}\right) _{c}-2b>\left( P_{0}\right) _{c}\sqrt{1-\frac{4b^{2}}{%
\alpha }}.  \label{xp}
\end{equation}%
If (\ref{lmax}) is satisfied, it is seen, with (\ref{ax}) taken into
account, that the left hand side of (\ref{xp}) is positive. Then, it is easy
to check that (\ref{xp}) is valid without any additional assumptions.

Thus we checked the validity of (\ref{Y}) in the point of collision for the
scenario under discussion. We want it to hold for any $r>r_{c}$ as well. If $%
\frac{d\omega _{+}}{dr}<0$, then $\frac{dY}{dr}>0$ that is sufficient for
the validity of (\ref{Y}) everywhere. For example, in the Kerr and
Kerr-Newman background $\frac{d\omega _{+}}{dr}<0$ for extremal black holes
and slightly overspinning metrics - e.g., see Fig. 1 and discussion in \cite%
{waldcol} and Fig. 2 in \cite{dc}.

In general, even if there exists a small interval where $\frac{d\omega _{+}}{%
dr}>0$, $\omega _{c}-\omega _{+}\approx -BN_{c}$ with the constant $B>0$ and 
$N_{c}\ll 1$, this only adds some inessential restriction on $l$ but does
not spoil the SPP. Indeed, in this interval we should require, according to
the positivity of (\ref{Y}), that%
\begin{equation}
L_{3}<\frac{\left( X_{3}\right) _{c}}{BN_{\max }}\text{,}
\end{equation}%
where $N_{\max \text{ }}$is the maximum value of $N$ within this interval.
Then, (\ref{Ll}) entails that 
\begin{equation}
l<\frac{\left( X_{3}\right) _{c}N_{c}}{BN_{\max }}\text{.}
\end{equation}

Thus, under rather weak assumptions that are satisfied in the Kerr and
Kerr-Newman cases, the SPP does exist. This includes also a wormhole
obtained by gluing two such metrics \cite{owh}.

Now, we can generalize (\ref{Ll}) and consider%
\begin{equation}
L_{3}\approx \frac{l}{N^{s}}\text{.}  \label{k}
\end{equation}

If $s>1$, this violates condition (\ref{m034}) and, hence, (\ref{d}) as
well. Therefore, this case should be rejected. Case $s=1$ is already
considered above. Let $s<1$. Then, $\Delta _{\pm }\approx \tilde{m}_{0}^{2}$%
, in (\ref{h3}) $H_{3.4}>0,$ so $\varepsilon _{3}=\varepsilon _{4}=+1$, and
only scenario 1 can be realized. We have from (\ref{x1})%
\begin{equation}
\left( X_{3}\right) _{c}=[\frac{\Delta _{+}}{2\tilde{m}_{0}^{2}}(X_{0}-P_{0}%
\sqrt{1-\frac{4\tilde{m}_{3}^{2}\tilde{m}_{0}^{2}}{\Delta _{+}^{2}}})]_{c}%
\text{,}
\end{equation}%
Neglecting in $P_{0}$ terms of the order $N^{2}$, we have%
\begin{equation}
X_{3}\approx X_{0}\frac{\tilde{m}_{3}^{2}}{\tilde{m}_{0}^{2}}\text{.}
\label{30}
\end{equation}%
This expression is finite. It is easy to check that in the point of
collision (\ref{Y}) is satisfied. As $\sigma _{3}=+1$ for scenario 1,
particle 3 moves after collision in the outward direction. If the assumption
about monotonic decrease of $\omega _{+}(r)$ (see above) is fulfilled, we
are faced with the SPP. Thus holds for any $0<s\leq 1$.

\section{Square of Energy in CM frame growing as $N_{c}^{-1}$}

In this section, we consider another asymptotic behavior, 
\begin{equation}
m_{0}^{2}\approx \frac{\beta }{N}\text{.}  \label{N}
\end{equation}

It is typical of the BSW \cite{ban} and Schnittman \cite{shnit} processes
that occur near the horizons of extremal black holes when critical particle
1 collides with usual particle 2. Our goal is to trace, how it happens that
the SPP is impossible. The details of the concrete energy bounds can be
found in \cite{j} - \cite{max}.

Using the Taylor expansion near the horizon%
\begin{equation}
\omega =\omega _{H}-B_{1}N+O(N^{2})\text{,}
\end{equation}%
one can infer \cite{j} that $X_{1}=B_{1}N_{c}L_{1}+O(N^{2})$, and%
\begin{equation}
\beta =2(X_{2})_{c}c\approx 2(X_{0})_{c}c\text{,}
\end{equation}%
where%
\begin{equation}
c=B_{1}L_{1}\pm \sqrt{B_{1}^{2}L_{1}^{2}-\tilde{m}_{1}^{2}}.
\end{equation}

We know already that for finite $L_{3}$ the SPP is impossible. We would like
to examine, what happens if, from the very beginning, we adjust a value of $%
L_{3}$ to the point of collision in such a way that $L_{3}$ is formally
divergent in the limit $N_{c}\rightarrow 0$. One of simplest choices is%
\begin{equation}
L_{3}\approx \frac{l}{\sqrt{N_{c}}}\text{.}  \label{root}
\end{equation}%
Then, it follows from (\ref{x1}), (\ref{x2}) that%
\begin{equation}
\left( X_{3}\right) _{c}\approx \frac{X_{0}}{2}\left( 1-\sqrt{1-\frac{4b^{2}%
}{\beta }}\right) \text{, }b\leq \frac{\sqrt{\beta }}{2}.
\end{equation}%
\begin{equation}
\left( X_{4}\right) _{c}\approx \frac{X_{0}}{2}\left( 1+\sqrt{1-\frac{4b^{2}%
}{\beta }}\right) \text{,}
\end{equation}%
notation $b$ from (\ref{b}) is used, $E_{3}=\omega _{c}L_{3}+\left(
X_{3}\right) _{c}$, so%
\begin{equation}
E_{3}\approx \frac{\omega _{H}l}{\sqrt{N_{c}}}\text{.}
\end{equation}%
Formally, we have large positive $E_{3}$ and large negative $E_{4}$. Now, $%
\Delta _{\pm }=O(N_{c}^{-1})$, so in (\ref{h3}) the negative term dominates
and $\varepsilon _{3}=\varepsilon _{4}=-1$. It is seen that scenario 3 or 5
can be realized.

Thus we have two usual particles that move towards a black hole and fall in
it since there are no turning points. In principle, this case (not discussed
in literature before)\ can be of some interest for more involved scenarios
that include consideration of events inside a black hole. However, for our
purposes, the case in question is completely useless since nothing comes out
to infinity and the SPP\ is impossible.

Instead of (\ref{root}), one can take%
\begin{equation}
L_{3}=\frac{l}{N_{c}^{\rho }}
\end{equation}%
with an arbitrary positive $\rho $. If $\rho >\frac{1}{2}$, condition (\ref%
{m034}) is also violated. Thus this case cannot be realized. Case $\rho =%
\frac{1}{2}$ is already considered above. Let now $\,0<\rho <\frac{1}{2}$.
Then, repeating consideration step by step, one can check that $\Delta _{\pm
}\approx \tilde{m}_{0}^{2}\approx \frac{\beta }{N_{c}}$, $\varepsilon
_{3}=\varepsilon _{4}=-1$, so only $\sigma _{3}=-1$ is possible as is seen
from (\ref{list}), we deal with scenario 3 or 5. In doing so, there is no
turning point, $P_{3}\approx X_{0}\frac{\tilde{m}_{3}^{2}}{\tilde{m}_{0}^{2}}%
\approx X_{0}\frac{b^{2}}{\beta }N^{1-2p},$ particle 3 approaches the
horizon and falls in a black hole. Therefore, the SPP\ is impossible
similarly to the case $\rho =\frac{1}{2}$.

To conclude this Section, we will discuss briefly also collisions near
nonextremal black holes. Although the BSW effect was found and remains more
popular just for extremal black holes, in somewhat modified form it exists
also for nonextremal ones. This was shown in \cite{gp} for the Kerr metric
and generalized to arbitrary axially symmetric stationary black holes in 
\cite{prd}. In contrast to the extremal case, for nonextremal one the
critical particle cannot reach the horizon and at first it would seem that
this fact makes the BSW effect impossible. However, it turned out that if
one takes a near-critical particle instead of the critical, with deviation
from the critical relation between the energy and angular momentum having
the order $N$ (see eq. 18 of \cite{gp} and eq. 18 in \cite{prd}, where $%
\varepsilon $ is implied to have the order $N$ with the restriction (18)
given in \cite{prd}), then the effect does take place with (\ref{N}) being
valid. Then, the consideration of this Section applies directly with the
same conclusion that the SPP\ is absent.

\section{Discussion and conclusions}

Thus we developed a scheme that enables us to classify all possible
scenarios of collisions in the equatorial plane. We found only 5
nonequivalent types of such scenarios. The approach works both for finite
and divergent energy in the CM frame and is applicable to diverse types of
space-times. Information about possibility or impossibility of the SPP is
encoded in the rate with which the energy in the CM frame diverges when $%
N_{c}\rightarrow 0$.

This allowed us to explain and unify previous observations. In doing so,
physical nature of the metric reveals itself indirectly, so qualitatively
different space-times and scenarios can give the same result. For example,
the dependence $m_{0}^{2}\sim N_{c}^{-2}$ can arise either in the two-step
scenario in the background of a naked singularity for slightly overspinned
metric or in collisions of two usual particles one of which comes from the
white hole horizon directly. The result is the same, in both cases the SPP
is possible. The results are summarized in Table 1.

\begin{tabular}{|p{1in}|p{1in}|p{1in}|p{1in}|p{1in}|p{1in}|}
\hline
$m_{0}^{2}$ & $\sigma _{1}\sigma _{2}$ & $N_{c}$ & Particles & Object & SPP
\\ \hline
finite & $\pm 1$ & $\sim 1$ & arbitrary & any & no \\ \hline
$N_{c}^{-2}$ & $-1$ (head-on) & $\ll 1$ & both usual & wormhole \cite{owh},
white hole \cite{gpw}, naked singularity \cite{waldcol} & yes \\ \hline
$N_{c}^{-1}$ & $\pm 1$ & $\ll 1$ & critical and usual & extremal black hole 
\cite{j} - \cite{frac} & no \\ \hline
$N_{c}^{-1}$ & $-1$ & $\ll 1$ & near-critical and usual & nonextremal black
hole \cite{gp}, \cite{prd} & no \\ \hline
\end{tabular}

Table 1. Possibility/impossibility of the super-Penrose process in different
cases.

We would like to stress that our conclusions are derived in a simple and
straightforward manner, directly in the original frame, we did not need to
pass into the centre of mass frame and back in contrast to \cite{inf}, \cite%
{waldcol}. We believe that the corresponding approach can be useful also in
other problems connected with high energy particle collisions in the region
of strong gravity. It is of interest to generalize it to the charged case
whose properties are expected to be very different from the neutral one.

\begin{acknowledgments}
The work is performed according to the Russian Government Program of
Competitive Growth of Kazan Federal University.
\end{acknowledgments}


\begin{thebibliography}{99}
\bibitem{ban} M. Ba\~{n}ados, J. Silk and S.M. West, Kerr Black Holes as
Particle Accelerators to Arbitrarily High Energy, Phys. Rev. Lett. \textbf{%
103} (2009) 111102 [arXiv:0909.0169].

\bibitem{pir1} T. Piran, J. Katz, and J. Shaham, High efficiency of the
Penrose mechanism for particle collision, Astrophys. J. \textbf{196}, L107
(1975).

\bibitem{pir2} T. Piran and J. Shaham, Production of gamma-ray bursts near
rapidly rotating accreting black holes, Astrophys. J. \textbf{214}, 268
(1977).

\bibitem{pir3} T. Piran and J. Shanam, Upper Bounds on Collisional Penrose
processes near rotating black hole horizons, Phys. Rev. D \textbf{16}, 1615
(1977

\bibitem{j} T. Harada, H. Nemoto and U. Miyamoto, Upper limits of particle
emission from high-energy collision and reaction near a maximally rotating
Kerr black hole, Phys. Rev\textit{. }D\textbf{\ 86,} (2012) 024027; \textbf{%
86,} (2012) 069902(E) [arXiv:1205.7088].

\bibitem{p} M. Bejger, T. Piran, M. Abramowicz, and F. H\aa kanson,
Collisional Penrose process near the horizon of extreme Kerr black holes,
Phys. Rev. Lett.\textbf{\ 109} (2012) 121101 [arXiv:1205.4350].

\bibitem{z} O. Zaslavskii, Energetics of particle collisions near dirty
rotating extremal black holes: Ba\~{n}ados-Silk-West effect versus Penrose
process, Phys. Rev. D\textbf{\ 86} (2012) 084030 [arXiv:1205.4410].

\bibitem{shnit} J. D. Schnittman, Revised upper limit to energy extraction
from a Kerr black hole, Phys.Rev.Lett. \textbf{113}, 261102 (2014),
[arXiv:1410.6446].

\bibitem{cons} T. Harada, K. Ogasawara, U. Miyamoto, Consistent analytic
approach to the efficiency of collisional Penrose process, Phys. Rev. D 
\textbf{94}, 024038 (2016) [arXiv:1606.08107].

\bibitem{pirmax} E. Leiderschneider and T. Piran, Maximal efficiency of the
collisional Penrose process, Phys. Rev. D 93, 043015 (2016)
[arXiv:1510.06764].

\bibitem{max} O. Zaslavskii, Maximum efficiency of the collisional Penrose
process, Phys. Rev. D \textbf{94}, 064048 (2016) [arXiv:1607.00651].

\bibitem{frac} O. B. Zaslavskii, Is the super-Penrose process possible near
black holes? Phys. Rev. D \textbf{93}, 024056 (2016), [arXiv:1511.07501].

\bibitem{card} E. Berti, R. Brito and V. Cardoso, Ultra-high-energy debris
from the collisional Penrose process, Phys. Rev. Lett. \textbf{114}, 251103
(2015), [arXiv:1410.8534].

\bibitem{mod} O. Zaslavskii, Unbounded energies of debris from head-on
particle collisions near black holes, Mod. Phys. Lett. A \textbf{30} (2015)
1550076, [arXiv:1411.0267].

\bibitem{esc} K. Ogasawara, T. Harada, U. Miyamoto, and T. Igata, Escape
probability of the super-Penrose process, Phys. Rev. D \textbf{95}, 124019
(2017) [arXiv:1609.03022].

\bibitem{gpw} A. Grib and Yu. V. Pavlov, Are black holes totally black?\
Grav. Cosmol. \textbf{21}, 13 (2015); [arXiv: 1410.5736].

\bibitem{pir15} E. Leiderschneider and T. Piran, Super-Penrose collisions
are inefficient - a Comment on: Black hole fireworks: ultra-high-energy
debris from super-Penrose collisions, arXiv:1501.01984.

\bibitem{epl} O. B. Zaslavskii, General limitations on trajectories suitable
for super-Penrose process, Europhys. Lett. \textbf{111} 50004 (2015),
[arXiv:1506.06527].

\bibitem{headon} O. B. Zaslavskii, Ultrahigh energy head-on collisions
without horizons or naked singularities: General approach. Phys. Rev. D 
\textbf{88}, 044030 (2013) [arXiv:1305.6136].

\bibitem{inf} M. Patil, T. Harada, Ken-ichi Nakao, P. S. Joshi and M.
Kimura, Infinite efficiency of the collisional Penrose process: Can a
overspinning Kerr geometry be the source of ultrahigh-energy cosmic rays and
neutrinos? Phys. Rev. D \textbf{93}, 104015 (2016) [arXiv:1510.08205].

\bibitem{waldcol} I. V. Tanatarov and O. B. Zaslavskii, Gen. Relativ. Gravit 
\textbf{49}, 119, Collisional super-Penrose process and Wald inequalities,
[arXiv:1611.05912].

\bibitem{owh} O.B. Zaslavskii, Super-Penrose process and rotating wormholes,
Phys. Rev.\textit{\ }D\textbf{\ 98} (2018) 104030 [arXiv:1807.11033].

\bibitem{rn} O. B. Zaslavskii, Energy extraction from extremal charged black
holes due to the BSW effect. Phys. Rev\textit{. }D\textbf{\ 86}, 124039
(2012) [arXiv:1207.5209].

\bibitem{prd} O.B. Zaslavskii, Acceleration of particles as universal
property of rotating black holes, Phys. Rev\textit{. }D\textbf{\ 82} (2010)
083004 [arXiv:1007.3678]

\bibitem{wald} Wald, R.M.: Energy limits on the Penrose process. Astrophys.
J. \textbf{191}, 231 (1974).

\bibitem{kras} S. Krasnikov, Schwarzschild-like wormholes as accelerators,
Phys. Rev. D \textbf{98}, 104048 (2018) [arXiv:1807.00890].

\bibitem{dc} D. Pugliese and H. Quevedo, Disclosing connections between
black holes and naked, singularities: horizon remnants, Killing throats and
bottlenecks, Eur. Phys. J. C \textbf{79}, 209 (2019) [arXiv:1902.07917].

\bibitem{gp} A.A. Grib and Yu.V. Pavlov, On particles collisions in the
vicinity of rotating black holes, Pis'ma v ZhETF \textbf{92}, 147 (2010)
[JETP\ Letters \textbf{92}, 125 (2010)].
\end{thebibliography}
\end{document}